\documentclass[%
 reprint,
showpacs,
showkeys,
 amsmath,
 amssymb,
 aps,
pre
]{revtex4-1}

\usepackage{graphicx}% Include figure files
\usepackage{dcolumn}% Align table columns on decimal point
\usepackage{bm}% bold math
\usepackage{hyperref}% add hypertext capabilities
\begin{document}

%----------------------------------------------------------------------------------------------

\title{Probing density waves in fluidized granular media with diffusing-wave spectroscopy}% Force line breaks with \\

\author{Philip Born}
\affiliation{Institut f\"ur Materialphysik im Weltraum, Deutsches Zentrum f\"ur Luft- und Raumfahrt (DLR), 51170 K\"oln, Germany }
 \email{Philip.Born@dlr.de}
\author{Steffen Reinhold}%
\affiliation{Institut f\"ur Materialphysik im Weltraum, Deutsches Zentrum f\"ur Luft- und Raumfahrt (DLR), 51170 K\"oln, Germany }
\author{Matthias Sperl}
\affiliation{Institut f\"ur Materialphysik im Weltraum, Deutsches Zentrum f\"ur Luft- und Raumfahrt (DLR), 51170 K\"oln, Germany }

\date{\today}% It is always \today, today,

%----------------------------------------------------------------------------------------------

\begin{abstract}
Density waves are characteristic for fluidized beds and affect measurements on liquid-like dynamics in fluidized granular media. Here, the intensity autocorrelation function as obtainable with diffusing-wave spectroscopy is derived in the presence of density waves. The predictions by the derived form of the IACF match experimental observations from a gas-fluidized bed. The model suggests separability of the contribution from density waves from the contribution by microscopic scatterer displacement to the decay of correlation, thus paves the way for characterizing microscopic particle motions using diffusing-wave spectroscopy as well as heterogeneities in fluidized granular media.
\end{abstract}

\maketitle

%\onehalfspacing

%--------------------------------------------------------------------------------------------------

\section{Introduction}
\label{Intro}

Characterization of the particle dynamics encountered in fluidized granular media, both the dynamics of the individual grains and the emergent regimes of fluidization, is crucial for advances in research on three-dimensional granular flows \cite{Durian2000}. Approaches to this challenge were made in the last two decades to track particle motion in fluidized beds using coherent laser light \cite{Durian1997a,Durian1997,Pak2001,Biggs2006,Goldman2006,Biggs2007}. These measurements rely on the principle of diffusing-wave spectroscopy (DWS), which allows connecting temporal intensity fluctuations of coherent light to displacements of microscopic scattering centers in opaque samples by the intensity autocorrelation function (IACF) \cite{Maret1990,Weitz1993}. The DWS measurements on fluidized granular media received some attention, as they supported analogies among dense granular media and thermal glassy systems \cite{Goldman2006,Biggs2007}.

For such an analogy to hold, granular media needs some form of agitation, such as the one present in a fluidized bed. However, basically all fluidized beds are unstable and exhibit particle number density waves that propagate along the flow direction of the fluid \cite{Jackson2000}. The wavelengths of these density fluctuations are on the order of the container size of the fluidized bed, and can be observed by pressure fluctuations or by incoherent optical probes \cite{Jackson1969,Gouesbet1995,Johnsson2000,Castellanos2003,Castellanos2005}. These measurements confirmed the presence of number density waves in fluidized beds at all levels of fluidization, even in the state commonly referred to as uniform fluidization \cite{Jackson2000,Castellanos2003,Jackson1968}.

The number density waves, which can be observed as intensity fluctuations when using the incoherent optical probes, should also leave a signature in the intensity fluctuations as observed in the DWS measurements using coherent light sources. We evaluate in the following section the consequences of density waves for the intensity autocorrelation function. Then, in Sec.~\ref{Exp}, we test the theoretical predictions by measurements in a gas-fluidized bed. The results show that the density waves become apparent in the IACF by a second decay with trailing oscillations, which signifies the periodicity of the density waves. The derived form of the IACF indicates that the two contributions to the intensity fluctuations, the density waves and the phase shifts created by microscopic motions of scattering centers, can be separated for correct interpretation.

%--------------------------------------------------------------------------------------------------

\section{Theoretical considerations}
\label{theo}

A central quantity obtainable in DWS measurements is the time-averaged intensity autocorrelation function (IACF) $\left<I(t)I(t+\tau)\right>$, where $I(t)$ is the intensity at some time $t$, $\tau$ denotes a delay time, and the brackets indicate temporal averaging. The temporal averaging is not necessarily equal to ensemble averaging, as granular media are not inherently ergodic \cite{John1989,Frenkel2012}, and the density waves considered here prevent a stationary state. We derive a formulation of the IACF in the presence of density waves within the frame of path-bound propagation of light. 

In highly opaque samples like granular media the incoming electromagnetic wave from a source is strongly scattered and eventually fades, and multiple waves propagate from scattering center to scattering center. This multiple scattering and eventually diffusive wave propagation can be represented by propagation of fields along distinct paths \cite{Weitz1993}. The total field $E(t)$ at the position of a detector becomes the sum of the waves which propagated along individual paths $P$:

\begin{equation}
    E(t) = \sum_{P}{E_{P} e^{i\Phi_{P}(t)}},
    \label{eq:Etstat}
\end{equation}

where $E_{p}$ represents the amplitudes and $\Phi_{P}(t)$ the phases of the individual fields. The field autocorrelation function is linked to phase shifts $\Delta\Phi_{P}(\tau)$ along the paths with time,

\begin{eqnarray}
    \left<E(t)E^{\ast}(t+\tau)\right> &=& \left<\left(\sum_{P}{E_{P} e^{i\Phi_{P}(t)}}\right) \right. \nonumber \\
    & & \cdot \left. \left(\sum_{P'}{E_{P'}^{\ast} e^{-i\Phi_{P'}(t+\tau)}}\right)\right> \nonumber \\
     &=& \sum_{P}{\left<|E_{P}|^{2}\right>\left<e^{i(\Phi_{P}(t)-\Phi_{P}(t+\tau)}\right>} \nonumber \\
     &=& \sum_{P}{\left<I_{P}\right>\left<e^{i\Delta\Phi_{P}(\tau)}\right>}.
    \label{eq:EACF}
\end{eqnarray}

Here the conventional assumptions were made, that phase and amplitude of the field of a certain path are uncorrelated, and that phases along different paths, $P\neq P'$, are uncorrelated (i.e., we assume that phases are evenly distributed over intervals of 2$\pi$), so that only terms with $P = P'$ contribute. The phase shifts of the waves are linked to the displacements of scattering centers and the length of the respective paths, and statistical considerations on the path length distribution then leads to the conventional scheme of DWS \cite{Weitz1993,Maret1990b}. 

Experimentally accessible are only intensities, $I(t) = E(t)\cdot E(t)^{\ast}$. The \textsc{Siegert}-relation establishes a connection between the intensity autocorrelation function and the field autocorrelation function \cite{Berne1976}:

\begin{equation}
    \left<I(t)I(t+\tau)\right> = \left< I(t)\right>^{2} + \left|\left<E(t)E^{\ast}(t+\tau)\right>\right|^{2}.
    \label{eq:Siegert}
\end{equation}

$E(t)$ and $E^{\ast}(t)$ are assumed normally distributed variables in the derivation of the \textsc{Siegert}-relation. This holds true for the total field in Eq.~(\ref{eq:Etstat}) by the central limit theorem, as $E(t)$ is a sum of fields with stationary uncorrelated amplitudes and evenly distributed random phases. A normal distribution of the field values in time results in an exponential distribution of intensity values, if instantaneous intensities could be measured, or to $\Gamma$-distributed intensity values if some time-integration is involved \cite{Goodman1985}. The measured IACFs are usually normalized by their long delay time limit $\left<I(t)I(t+\tau)\right>|_{\tau\rightarrow\infty} = \left<I(t)\right>\left<I(t+\tau)\right> = \left<I(t)\right>^{2}$, as is done in the hardware correlation used in the experiments. This leads to an intercept ($\tau\rightarrow0$) of the IACF of 2 within the validity of the \textsc{Siegert}-relation:

\begin{eqnarray}
    \left.\frac{\left<I(t)I(t+\tau)\right>}{\left<I(t)\right>^{2}}\right|_{\tau\rightarrow0} &=& 1 + \left.\frac{\left|\left<E(t)E^{\ast}(t+\tau)\right>\right|^{2}}{\left<I(t)\right>^{2}}\right|_{\tau\rightarrow0} \nonumber \\
    &=& 2 .
    \label{eq:Siegert2}
\end{eqnarray}

\begin{figure}
    \centering
    \includegraphics[width=0.44\textwidth]{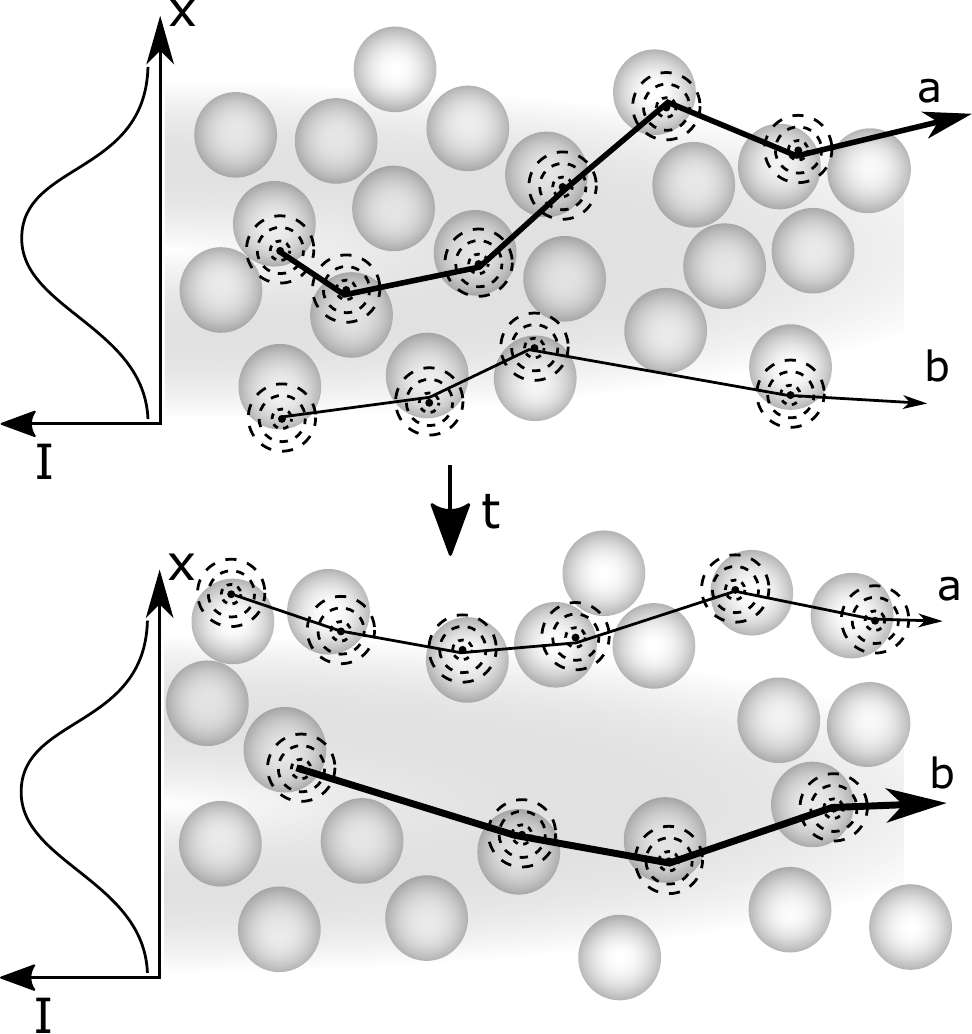}
    \caption{Schematic drawing of the time-dependency of path amplitudes in a fluidized sample, with two highlighted paths a and b that change between large and small transported amplitude with time. A laser beam with radially decaying intensity profile I(x) illuminates a sample with microscopic scattering centers on the surface or within the bulk of macroscopic particles. The position of the scattering centers relative to each other and relative to the incoming beam changes with rotation or translation of the particles. The paths will persist due to the continuity of the angular scattering at the microscopic scattering centers, but the transported amplitude will become position-dependent.}
    \label{fig:1}
\end{figure}

The exact nature of the scattering centers linked to the paths of diffusive wave propagation is hard to define in an ensemble of large granular particles. Certainly, Mie-like scattering of the large spheres play a role, but also scattering from surface asperities, scattering from inhomogeneities within the particle material and maybe even diffraction within narrow gaps formed by the densely packed particles can be non-negligible. Thus paths can be formed by scattering centers within or on the surface of the particles and change shape when the particle translate or rotate (Fig.~\ref{fig:1}). 

Density waves create periodic fluctuations of the concentration of particles in certain sub-volumes of the sample, in particular also in the region illuminated by the laser beam.  Inhomogeneous and collimated light sources like laser beams with Gaussian intensity distributions will thus be much more sensitive to this fluctuation mechanism than extended, homogeneous light sources, which may average over the wavelength of the density wave. The effect of a density fluctuation in the illuminated region could be taken into account by changing the ensemble of paths $P$, over which is summed in the previous equations, and consequently to a time-dependent path-length distribution, with complications to the DWS evaluation. Here we suggest to take the density waves into account in a different way: Electromagnetic waves extend infinitely, even the intensity of a Gaussian laser beam rapidly decays radially, but has no strict cut-off radius. Also the scattering at the various scattering centers has a non-vanishing scattering amplitude at any angle. This motivates the view, that irrespective how the particles associated to a respective path move relative to each other and relative to the light source, the path will still exist and will be excited by the incoming light. Only the intensity transported along the path may become insignificant with displacement of the particles (Fig.~\ref{fig:1}). We thus assume a stationary ensemble of paths, and only the amplitude of the electromagnetic wave propagating along a respective path is changing in time:

\begin{equation}
    E(t) = \sum_{P}{E_{P}(t) e^{i\Phi_{P}(t)}}.
    \label{eq:Et}
\end{equation}

It may be hypothesized that fewer and shorter paths carry a large amplitude during a low-density state of the volume illuminated by the laser, while during high-density states more and longer paths carry an overall lower amplitude. The field amplitudes in equation~(\ref{eq:Et}) will follow the temporal behavior of the density waves in the fluidized bed, thus in general they will not be uncorrelated anymore. This violates the assumptions for the central limit theorem, and limits the applicability of the \textsc{Siegert}-relation. A deviation from a normal distribution of the fields could be tested experimentally by comparing the distribution of the intensity values to a Gamma-distribution  \cite{Goodman1985}.

We calculate the instantaneous intensity $I(t) = E(t)\cdot E(t)^{\ast}$, the time-averaged intensity $\left<I(t)\right>$ and the time-averaged intensity autocorrelation $\left<I(t)I(t+\tau)\right>$ from this total electric field to clarify differences to the case without density waves:

\begin{equation}
    I(t) = \sum_{P}{\sum_{P'}{E_{P}(t)E_{P'}^{\ast}(t) e^{i(\Phi_{P}(t)-\Phi_{P'}(t))}}}.
    \label{eq:It}
\end{equation}

We again use the assumptions, that phase and amplitude of the field of a certain path and the phases along different paths $P\neq P'$ are uncorrelated, to obtain the averaged intensity:

\begin{eqnarray}
    \left<I(t)\right> &=& \left<E(t)\cdot E^{\ast}(t)\right> \nonumber \\
     &=& \left<\sum_{P}{\sum_{P'}{E_{P}(t)E_{P'}^{\ast}(t) e^{i(\Phi_{P}(t)-\Phi_{P'}(t))}}}\right> \nonumber \\
   &=& \sum_{P}{\sum_{P'}{\left<E_{P}(t)E_{P'}^{\ast}(t)\right> \left<e^{i(\Phi_{P}(t)-\Phi_{P'}(t))}\right>}} \nonumber \\
      &=& \sum_{P}{\left<|E_{P}(t)|^{2}\right>} \nonumber \\
      &=& \sum_{P}{\left<I_{P}(t)\right>} \nonumber \\
      &\equiv& \left<I_{t}(t)\right>. 
    \label{eq:Iave}
\end{eqnarray}

Here, we derived a time-dependent instantaneous total intensity $I_{t}(t)$, which is the summed up instantaneous intensity of all paths. The calculation of the full form of the IACF is then straight forward:

\begin{widetext}

\begin{eqnarray}
    \left<I(t)I(t+\tau)\right> &=& \left<E(t) E^{\ast}(t) \cdot E(t+\tau) E^{\ast}(t+\tau)\right> \nonumber \\
     &=& \left<\sum_{P}{\sum_{P'}{\sum_{P''}{\sum_{P'''} {E_{P}(t)E_{P'}^{\ast}(t)E_{P''}(t+\tau)E_{P'''}^{\ast}(t+\tau) e^{i(\Phi_{P}(t)-\Phi_{P'}(t)+\Phi_{P''}(t+\tau)-\Phi_{P'''}(t+\tau))}}}}}\right>.
    \label{eq:IACF}
\end{eqnarray}

\end{widetext}

As before, we separate the phase and amplitude averages and assume uncorrelated paths. Then
contributions arise only for $P = P'$, $P'' = P'''$, and $P = P''' \neq P' = P''$, so that the
intensity autocorrelation function is given by:

\begin{widetext}

\begin{eqnarray}
     &=& \left<\sum_{P}\sum_{P''}{E_{P}(t)E_{P}^{\ast}(t)E_{P''}(t+\tau)E_{P''}^{\ast}(t+\tau)}\right> \nonumber \\
     & & + \left<\sum_{P}{\sum_{P'}{E_{P}(t)E_{P}^{\ast}(t+\tau)E_{P'}(t)E_{P'}^{\ast}(t+\tau) e^{i(\Phi_{P}(t)-\Phi_{P}(t+\tau)-\Phi_{P'}(t)+\Phi_{P'}(t+\tau))}}}\right> \nonumber \\
     &=& \left<\sum_{P}\sum_{P''}{I_{P}(t)I_{P''}(t+\tau)}\right> \nonumber \\
     & & +  \left<\sum_{P}{\sum_{P'}{E_{P}(t)E_{P}^{\ast}(t+\tau)E_{P'}(t)E_{P'}^{\ast}(t+\tau) e^{i(\Phi_{P}(t)-\Phi_{P}(t+\tau))} e^{-i(\Phi_{P'}(t)-\Phi_{P'}(t+\tau))} }}\right>.
    \label{eq:IACF2}
\end{eqnarray}

\end{widetext}

Then we assume that the phase shifts fluctuate much more rapidly than the path amplitudes and thus can replace the amplitude terms in the second summand by their initial value. Using the notation above for the time-dependent total intensity and the independence of amplitude and phase we obtain:

\begin{widetext}

\begin{eqnarray}
   &=& \left<I_{t}(t)I_{t}(t+\tau)\right> \nonumber \\
     & & +  \left<\sum_{P}{\sum_{P'}{E_{P}(t)E_{P}^{\ast}(t)E_{P'}(t)E_{P'}^{\ast}(t) e^{i(\Phi_{P}(t)-\Phi_{P}(t+\tau))} e^{-i(\Phi_{P'}(t)-\Phi_{P'}(t+\tau))} }}\right> \nonumber \\
     &=& \left<I_{t}(t)I_{t}(t+\tau)\right> \nonumber \\
     & & +  \sum_{P}{\sum_{P'}{\left<I_{P}(t)I_{P'}(t)\right> \left< e^{i(\Phi_{P}(t)-\Phi_{P}(t+\tau))}\right> \left< e^{-i(\Phi_{P'}(t)-\Phi_{P'}(t+\tau))} \right> }} \nonumber \\
     &=& \left<I_{t}(t)I_{t}(t+\tau)\right> \nonumber \\
     & & +  \sum_{P}{\sum_{P'}{\left<I_{P}(t)I_{P'}(t)\right> }} \cdot \left|\left<e^{i\Delta\Phi_p(\tau)}\right>\right|^{2}.
    \label{eq:IACF3}
\end{eqnarray}

\end{widetext}

The last step could be considered as a reformulation of the \textsc{Siegert}-equation in the presence of time-dependent amplitudes (compare equations~(\ref{eq:EACF}) and~(\ref{eq:Siegert})):

\begin{widetext}

\begin{equation}
    \left<I(t)I(t+\tau)\right> = \left<I_{t}(t)I_{t}(t+\tau)\right> + \left|\left<E(t)E^{\ast}(t+\tau)\right>\right|^{2}.
    \label{eq:IACF4}
\end{equation}

\end{widetext}

The IACF thus has turned into the sum of two $\tau$-dependent contributions in the presence of time-dependent amplitudes, the phase shifts of the waves propagating along the paths and the fluctuations of the total instantaneous intensity. The intercept of the IACF normalized in the conventional way (cf. Eq.~(\ref{eq:Siegert2})) by the long time limit $\left<I(t)\right>^{2}$ depends on the fluctuations of the instantaneous intensity and in general will exceed 2:

\begin{eqnarray}
    \left.\frac{\left<I(t)I(t+\tau)\right>}{\left<I(t)\right>^{2}}\right|_{\tau\rightarrow0} &=& \left.\frac{\left<I_{t}(t)I_{t}(t+\tau)\right>}{\left<I(t)\right>^{2}}\right|_{\tau\rightarrow0} \nonumber \\
    & & + \left.\frac{\left|\left<E(t)E^{\ast}(t+\tau)\right>\right|^{2}}{\left<I(t)\right>^{2}}\right|_{\tau\rightarrow0} \nonumber \\
    &=&
    \frac{\left<I_{t}(t)^{2}\right>}{\left<I(t)\right>^{2}} + 1 \nonumber \\
    &\geq & 2 .
    \label{eq:Siegert3}
\end{eqnarray}

Summarizing, several import consequences follow for experiments using fluidized beds and DWS from the considerations and the derivation above. The electric field $E(t)$ at the detector will not be normally distributed, and consequently the intensity $I(t)$ will not be $\Gamma$-distributed anymore. Thus, also the values of the IACF and the field autocorrelation function at $\tau = 0$ will deviate from 2 and 1, respectively, the values derived for normally distributed fields \cite{Durian1999}. Most important, the IACF will exhibit decays of two distinct contributions. One arises from intensity fluctuations that follow the temporal behavior of density waves in the fluidized bed. The other contribution arises from phase shifts of the individual fields propagating along different paths, which carry information on microscopic displacements in the sample. Both contributions can be separated by subtracting one of the terms in Eq.~(\ref{eq:IACF4}). These predictions will be compared to results from a gas fluidized bed in the following.

%--------------------------------------------------------------------------------------------------

\section{Experimental results}
\label{Exp}

The experimental setup consists of a conventional gas-fluidized bed. Dry nitrogen is passed from below through a packing of opaque white 220~$\mu$m polystyrene particles resting on a glass frit in a glass tube with 10~mm inner diameter. The gas flow is controlled by a manual volume flow controller (Swaqelok), which allows adjusting volume flows between 0.5~l/h and 5~l/h with a resolution around 0.2~l/h. A Coherent Verdi G5 SLM (5~W, 532~nm, operated at 250~mW output power) is used as a light source. Light is detected in transmission through a linear polarizing filter (Owis GmbH) by a single mode fiber (Thorlabs), fed into a beamsplitter (Sch\"after+Kirchhoff) and finally guided into two avalanche photodiodes (ID Quantique). The signal is evaluated using a hardware correlator (ALV 7002/USB-25) by cross-correlation of the two detector signals to suppress afterpulsing effects. The hardware correlator provides a fast count rate trace with a time resolution of 200~$\mu$s and the IACF with 25~ns sampling time.

\begin{figure}
    \centering
    \includegraphics[width=0.44\textwidth]{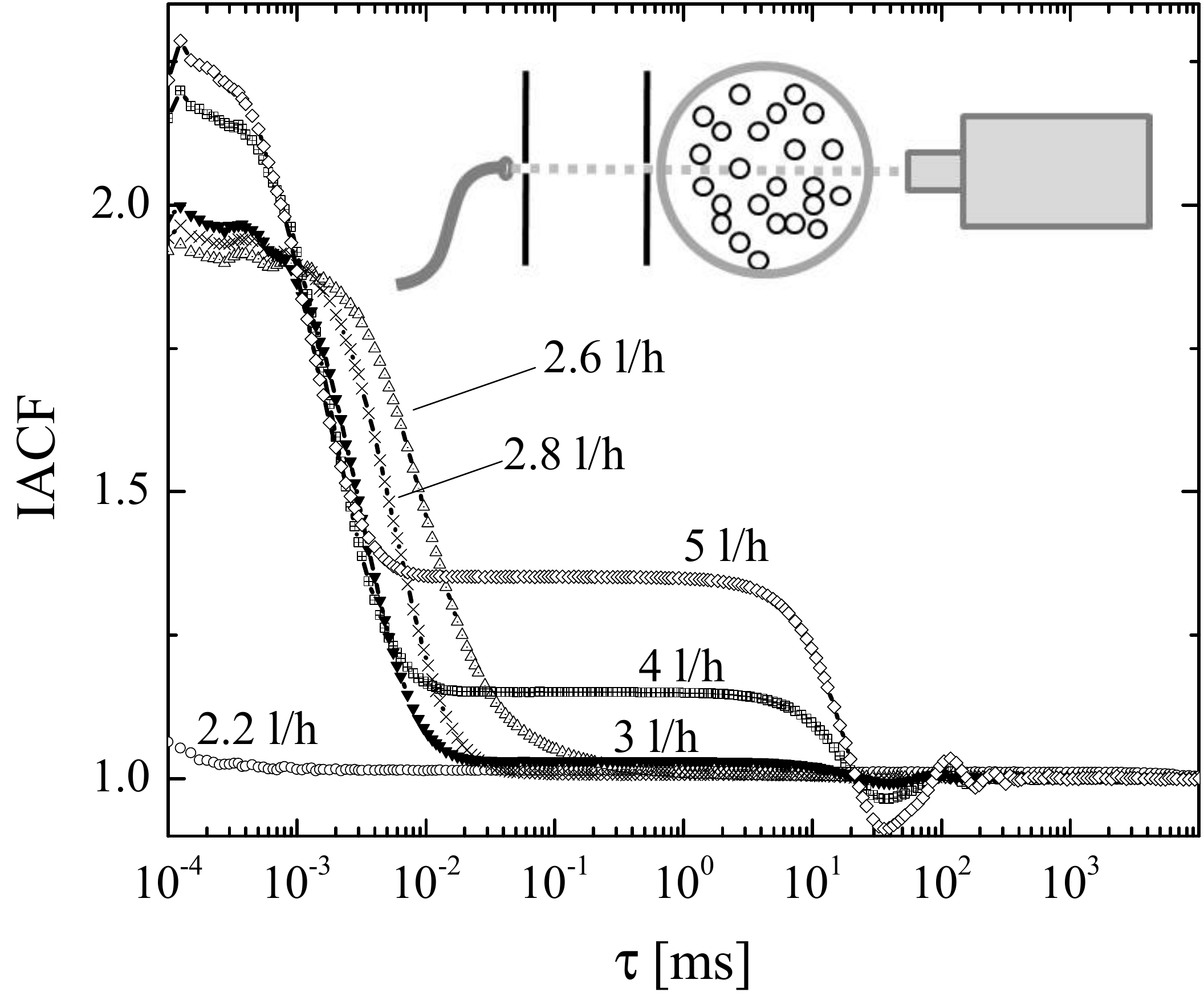}
    \caption{Normalized intensity autocorrelation functions (IACF) obtained from a fluidized bed at increasing gas flows. The inset shows a scheme of the setup, with laser, cylindrical fluidized bed, collimation and polarization filter. The labels indicate the gas flows for the respective curve. The IACFs exhibit two distinct decays and a growing intercept upon increased gas agitation.}
    \label{fig:2}
\end{figure}

The measured IACFs indicate that the sample stays static at low gas flows (below approximately 2.5~l/h, see Fig.~\ref{fig:2}). Then a single decay emerges in the IACF in a narrow regime of gas flows below 3~l/h. Above 3~l/h, two decays in the correlation functions emerge. The height of the second decay grows with increasing the gas flow. Simultaneously the intercept of the correlation curves rises above two. We note that the exact gas flow values at which the sample fluidized vary from one experimental run to another, which might be due to charging of the particles in the dry nitrogen stream.

\begin{figure}
    \centering
    \includegraphics[width=0.45\textwidth]{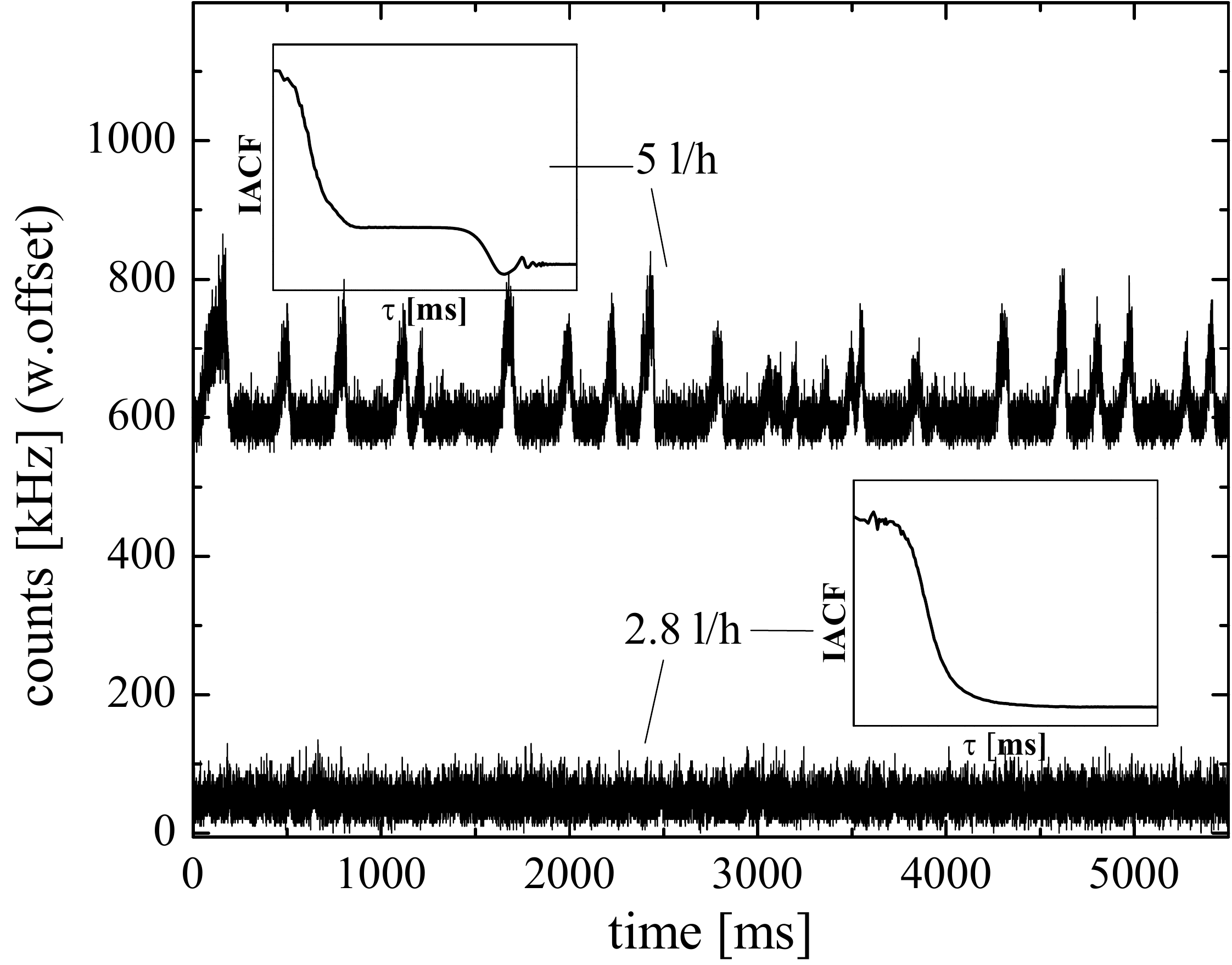}
    \caption{Intensity traces (photon counts/second) of an experiment with a single decay in the IACF and of an experiment with a double decay in the IACF (the latter trace is offset for clarity). A double-decay in the IACF is linked to periodic spikes in the intensity traces, showing fluctuations in the transmitted intensity.}
    \label{fig:3}
\end{figure}

The fast count rate traces of measurements with two decays in the IACF are qualitatively different
from measurements with a single decay (Fig.~\ref{fig:3}). Measurements with low gas flows and
single decay in the IACF exhibit count rates fluctuating randomly in a narrow band of intensity values. Measurements with high gas flows and two decays periodically exhibit broad spikes with increased intensity superimposed on top of the  the random fluctuations.

\begin{figure}
    \centering
    \includegraphics[width=0.45\textwidth]{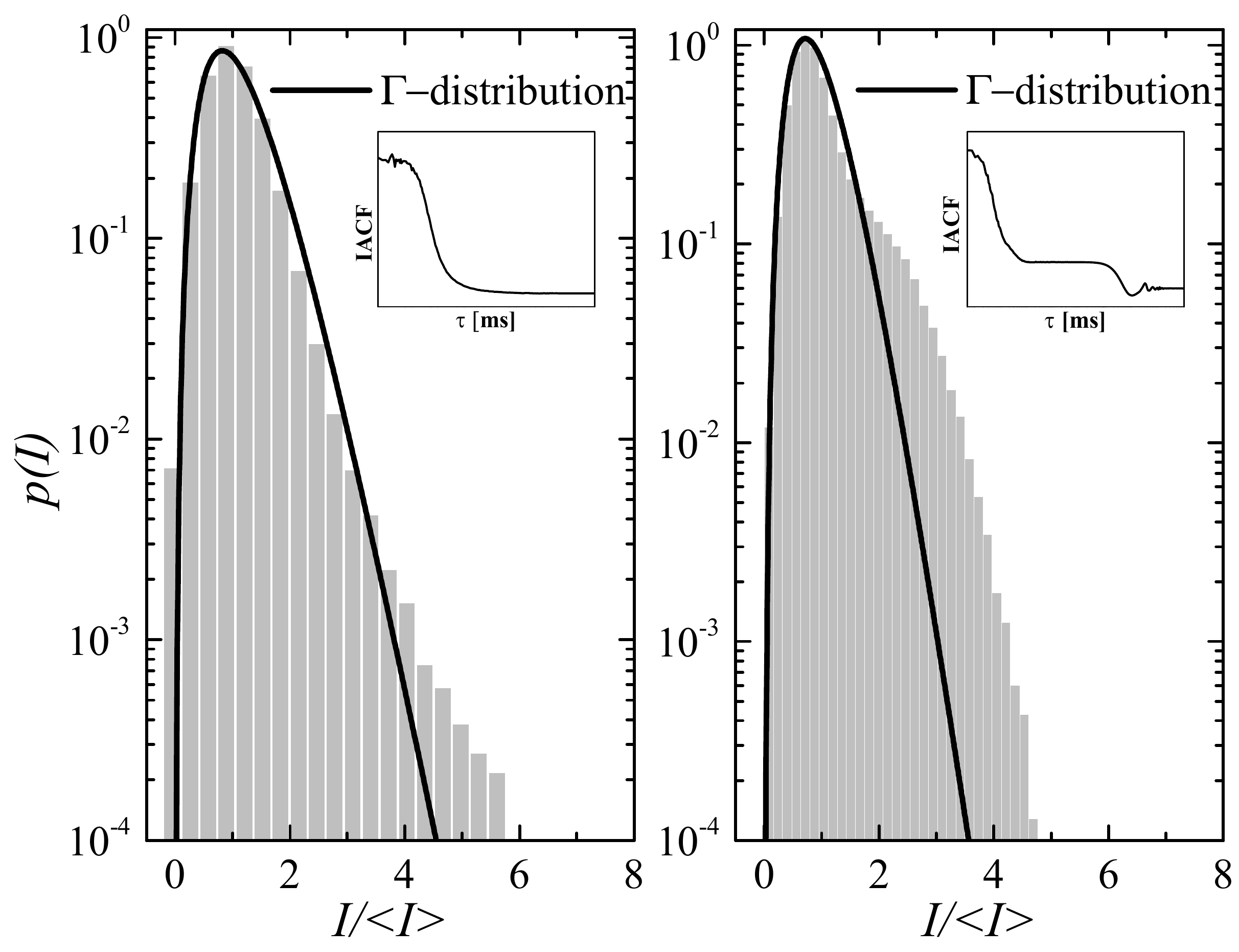}
    \caption{Time-averaged intensity probability distribution $p(I)$ for experiments with single- and double-decay in the IACF. The solid line is a fitted $\Gamma$-distribution, i.e. expectation from normally distributed intensity statistics. A broadening of the distribution compared to the expectation can be observed, which becomes much stronger for the measurement exhibiting a strong second decay.}
    \label{fig:4}
\end{figure}
\begin{figure}
    \centering
    \includegraphics[width=0.45\textwidth]{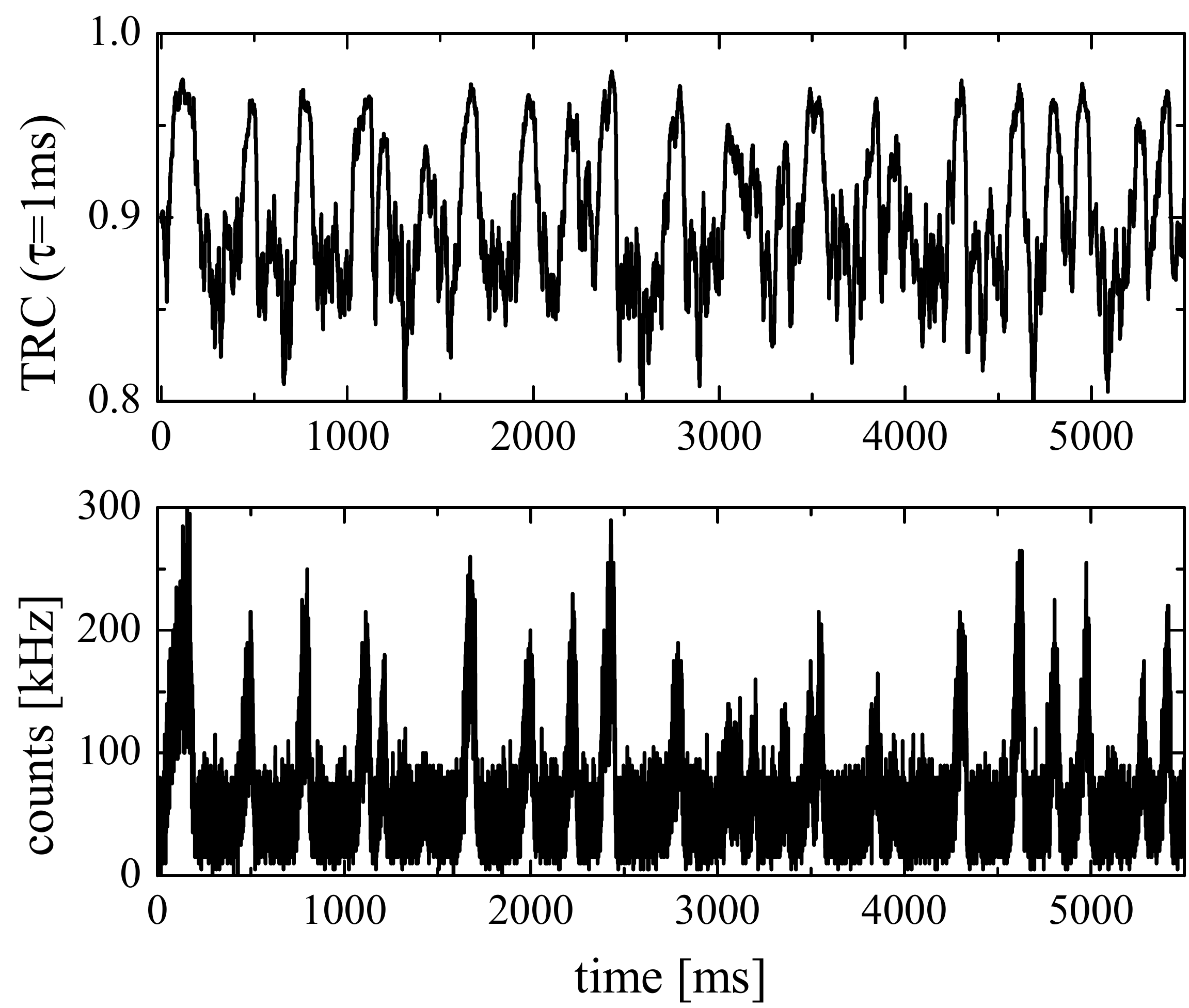}
        \caption{Results of the time resolved correlation (TRC) analysis of an sample with double-decay in the IACF. The upper graph gives the TRC traces with fixed 1ms delay time of an experiment with two decays, the lower graph the associated intensity trace. Periods with high transmitted intensity correlate with periods of conserved correlation.}
        \label{fig:5}
\end{figure}
The periodic spikes in intensity strongly affect the distribution of intensity values (Fig.~\ref{fig:4}). The distribution of intensity values is plotted together with a fitted $\Gamma$-distribution. The hardware correlator integrates intensity fluctuations over 200~$\mu$s to obtain the fast count rate traces. The intensity distribution from normally distributed intensities turn from an exponential distribution to a $\Gamma$-distribution with finite integration times, where the variance of the distribution is determined by the number of independent correlation intervals in the time integration interval \cite{Goodman1985}. The fit of a $\Gamma$--distribution to the normalized experimental intensity distributions gave 4.1 correlation intervals for the experiments with a single decay (2.6~l/h) and 4.9 correlation intervals for the experiment with two decays (5~l/h) within the integration time of 200~$\mu$s. This indicates that the fields become uncorrelated to a large extent after 50~$\mu$s for the lower gas flow, and within $\approx$40~$\mu$s for the sample with higher gas flow, what matches the observed first decay of the correlation functions (Fig.~\ref{fig:2}). A deviation of the intensity distribution from the expected $\Gamma$-distribution becomes apparent, which becomes enhanced for larger gas flows and measurements with pronounced second decay. 

We additionally calculate time resolved correlation (TRC) functions from the count rate traces to gain insight into the dynamics during periods of high and low transmissivity \cite{Cipelletti2003}:
\begin{equation}
TRC(t) = \frac{\left<I(t)\cdot I(t+\tau)\right>_{\Delta t}}{\left<1/2\cdot(I(t)^{2}+I(t+\tau)^{2})\right>_{\Delta t}}
\label{eq:}
\end{equation}
We take a moving average over a time interval of $\Delta t=100$~ms to take into account that we cannot average over many independent correlation areas as with a CCD camera. The delay time $\tau$ is set to 1~ms. At this time the fluctuations leading to the first decay in the IACF are readily averaged, but the fluctuations of the second decay should be well characterized. 

The time resolved correlation shows oscillations that follow the same periodicity as the intensity trace (Fig.~\ref{fig:5}). Spikes in the intensity traces correspond to periods where correlation is conserved most. This might indicate that during periods with higher density more longer and thus faster fluctuating paths contribute, while during periods with low density and high transmission shorter paths prevail.

%--------------------------------------------------------------------------------------------------

\section{Discussion}

The experimental observations support the considerations above for fluidized beds exhibiting density waves. The intensity autocorrelation functions exhibit two distinct drops in correlation, with an intercept exceeding 2. The intensity distribution obtained from fluidized bed measurements deviates from the prediction by normally distributed electric fields. The intensity and the time resolved correlation show periodic fluctuations in intensity and correlation. 

\begin{figure}
    \centering
    \includegraphics[width=0.45\textwidth]{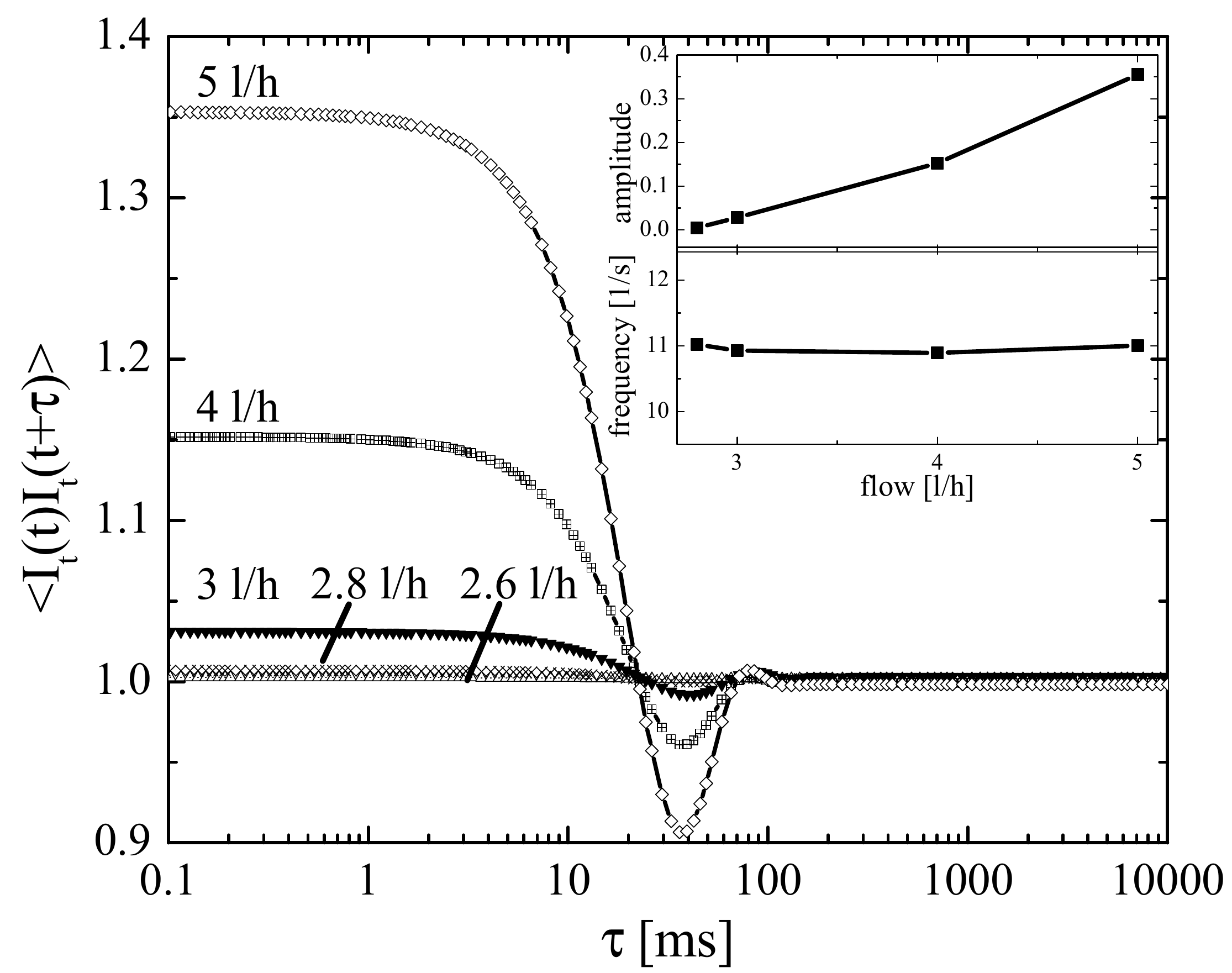}
        \caption{The isolated second decay of the experimental IACFs, exhibiting a remarkable constance in the time scales of the function. The inset gives the fit parameter amplitude and frequency of the oscillations in the second decay as a function of flow rate. The frequency stays constant around 11~Hz.}
        \label{fig:6}
\end{figure}

The derived equation for the IACF, Eq.~(\ref{eq:IACF4}), suggests the separability of the contributions from path amplitude fluctuations and from phase shifts. We fit a cosine function multiplied with an exponential decay to the second decay of IACF, in order to take the periodicity of the density waves with noise into account. This functional form fits the second decay with a coefficient of determination ($R^{2}$) of 0.998. The isolated contribution of the path amplitude fluctuations are given in Fig.~\ref{fig:6}. The amplitude of this fluctuations increase with increasing gas flow, while the frequency of these fluctuations stays remarkably constant over the whole range of gas flows at 11~Hz. This conserved time scale is in agreement with predictions of a dominant wave vector of the density fluctuations in fluidized beds \cite{Jackson2000}, and observations of strong periodicity in density fluctuations in deep fluidized beds by incoherent probes \cite{Johnsson2000}.

\begin{figure}
    \centering
    \includegraphics[width=0.45\textwidth]{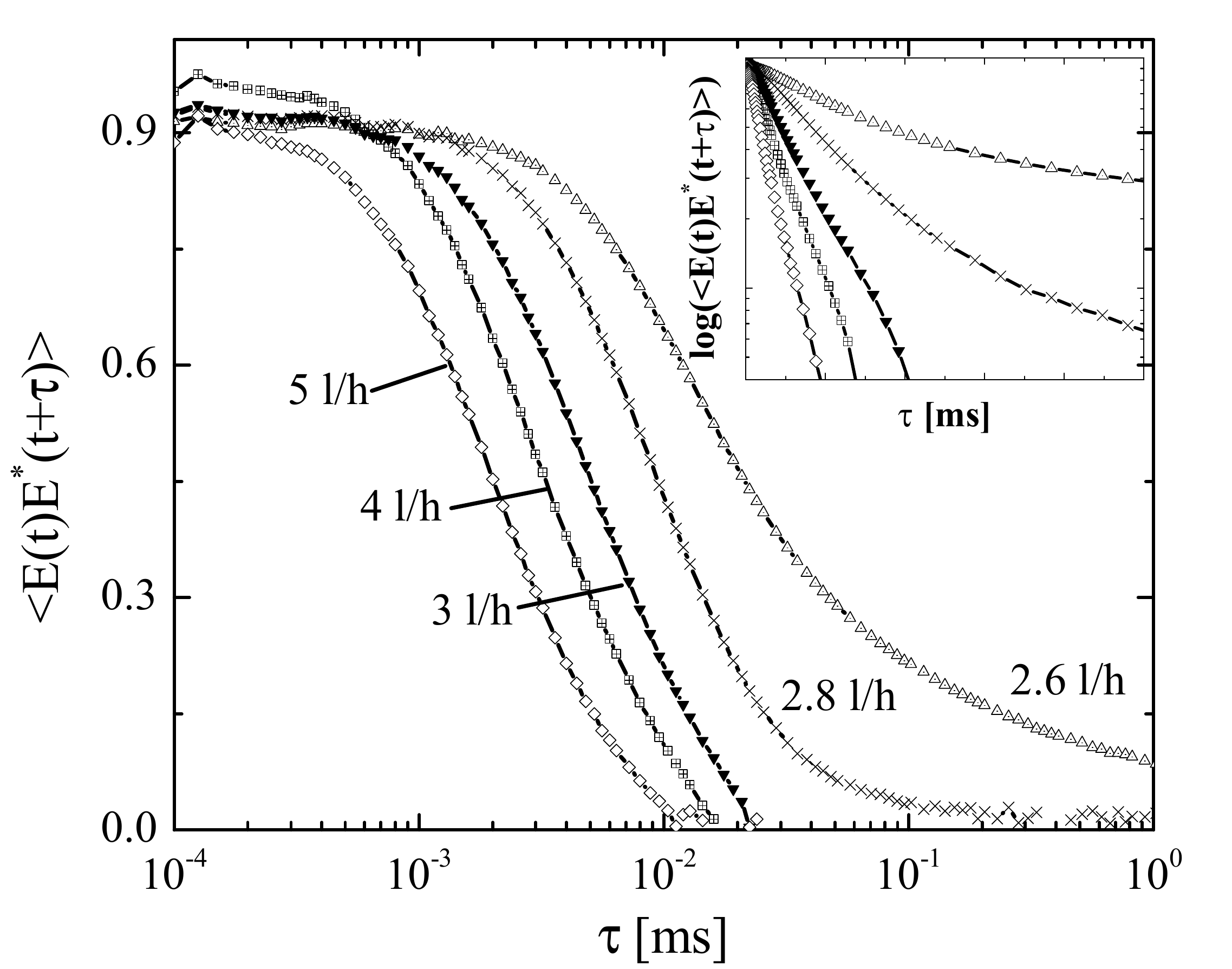}
        \caption{The isolated field autocorrelation functions after correction for the intensity fluctuations. The amplitude of the field autocorrelation stays close to 1 for all gas flows, showing the Gaussian statistics of the underlying phase fluctuations. The inset gives a log-lin-plot of the field correlations with symbols identical to the main panel. They turn from a stretched exponential to an exponential function with increased gas agitation.}
        \label{fig:7}
\end{figure}
The electric field autocorrelation functions are isolated after subtracting the fits to the second decay from the experimental IACFs and taking the square root (Fig.~\ref{fig:7}). A constant intercept of the correlation functions is recovered after correction for the amplitude fluctuations. The field autocorrelations decay faster with increasing the gas flow, indicating faster motion of microscopic scattering centers. Interestingly, the decay of the field autocorrelation function turns from a stretched exponential decay to an exponential decay, as can be seen in the inset of Fig.~\ref{fig:7}. This might be attributed to a transition from a sub-diffusive to diffusive motion of the scattering centers \cite{Maret1987}. Alternatively, averaging over many localized intermittent rearrangements on scales larger than the wavelength also results in an nearly exponential decay \cite{Durian1991}. However, the interpretation of these field autocorrelations has to be done with care. The curves are obtained by time-averaging over the non-stationary low-density and high-density states of the probed volume and the sample may exhibit dynamic heterogeneities, thus ensemble averaging might require additional efforts. Also, a conclusive interpretation of the field correlation function obtained from granular samples requires certainly further investigations, potentially including contributions from rotation of rough, inhomogeneous particles.

The interpretation and evaluation of an IACF is never non-ambiguous and requires additional assumptions about the investigated system. It is thus worth checking the plausibility of the interpretation of the particular shape of the measured IACFs presented here (sum of path amplitude fluctuations by density waves and phase shifts by microscopic displacements) by comparing to other possible interpretations.

The IACF derived in Sec.~\ref{theo} is very similar to the functions derived in the case of
source fluctuations \cite{Durian1999}, intermittency \cite{Durian2001}, and number fluctuations \cite{Pusey1979}. The measured IACFs (Fig.~\ref{fig:2}) alone hardly allow for a discrimination of those cases. The measured IACFs even would allow for the additional interpretation of increasingly glass-like localized particle dynamics upon increased gas flow, similar to the glassy interpretation in other experiments \cite{Goldman2006,Biggs2007}. These cases, however, make slightly different predictions and seem unlikely here:

Including \emph{source fluctuations} into the derivation of the IACF leads to predictions very similar to the path amplitude fluctuations introduced here. An additional time scale not related to scatterer dynamics will be present in the IACF and the probability distribution $p(I)$ will be wider than from normally distributed fluctuations alone. The IACF in this case becomes the product of two terms representing phase and amplitude fluctuations \cite{Durian1999}, not the sum as in our case. Such a contribution of a fluctuating source is unlikely in our case, as it should be present in all measurements independent of the gas flow.

\emph{Intermittent dynamics} lead to an IACF with additional terms added to represent the different dynamical states contributing to the total decay of correlation \cite{Durian2001}. A time-resolved correlation function allows quantifying the switching between the states \cite{Cipelletti2003}. Thus the calculated TRC-traces (Fig.~\ref{fig:5}) are in agreement with an interpretation of the IACFs as a result as intermittent dynamics. However, for intermittent dynamics the fields obey Gaussian statistics in all the dynamical states, and the distribution of intensity values $p(I)$ should not be altered (as in Fig.~\ref{fig:4}), and the amplitude of the IACF does not exceed the Gaussian prediction of 2 \cite{Durian2001}, as they do here (Fig.~\ref{fig:2}).

Microscopic \emph{localization} of particle dynamics does not include any modification to the intercept of the IACF and the distribution $p(I)$ nor any periodicity in the signal. Glassy localization thus cannot explain the intensity distribution (Fig.~\ref{fig:4}), the TRC-results (Fig.~\ref{fig:5}), and the observed periodicity in the second decay (Fig.~\ref{fig:6}), in addition to being counterintuitive to emerge upon increased gas flow.

\emph{Particle number fluctuations} actually lead to a prediction for the IACF and $p(I)$ that has a form very similar to the derived form here \cite{Pusey1979}. However, a proper sensitivity to number fluctuations can only be obtained in the regime of single scattering, not in the regime of multiple scattering, and number fluctuations do not lead to the observed periodicity in the second decay (as in Fig.~\ref{fig:6}).

Density waves are very common and basically happen in every stage of fluidization, not only for bubbling beds, and even for small particle sizes and water fluidized beds \cite{Jackson2000,Castellanos2003,Jackson1968}. This suggests that the data evaluation scheme presented here might be regularly considered when performing DWS measurements on fluidized beds.

%--------------------------------------------------------------------------------------------------

\section{Conclusion}

We derive a formulation of the intensity autocorrelation in the presence of density waves. The predictions match well the experimental observations obtained from a gas fluidized bed with granular particles. The model has similarities to source fluctuations, intermittency and number fluctuations, but can be discriminated by detailed inspection of the count rate traces by the intensity probability distribution and the time-resolved correlation.

The derived formulation allows separating the contribution from microscopic displacements of scattering centers, which result in phase shifts of electric fields transported through the sample, from the contribution by density waves, which result in fluctuations of the total instantaneous intensity. The microscopic motion becomes monotonically faster with increasing gas flow, while the density waves only increase their amplitude, and not their frequency.

The approach presented here should pave the way for exact characterization of particle displacements in fluidized beds, but might also support establishing methods to characterize the emergence of heterogeneity in granular media, with prospective applications to cooling, clustering, agglomeration and instabilities in fluidized beds.

%--------------------------------------------------------------------------------------------------

\begin{acknowledgments}

The authors thank Till Kranz for reviewing the manuscript. P. B. thanks Andreas Meyer for his continued support of the project. Financial support by DFG research unit FOR 1394 is gratefully acknowledged. 

\end{acknowledgments}

%--------------------------------------------------------------------------------------------------


\begin{thebibliography}{99}

\bibitem{Durian2000} D.~J. Durian, J. Phys. Condens. Matter \textbf{12}, 8A (2000).

\bibitem{Durian1997a} N. Menon and D. J. Durian, Science \textbf{275}, 5308 (1997).

\bibitem{Durian1997} N. Menon and D. J. Durian, Phys. Rev. Lett. \textbf{79}, 18 (1997).

\bibitem{Pak2001} S. Y. You and H. K. Pak, Journal-Korean Phys. Soc. \textbf{38}, 5 (2001).

\bibitem{Biggs2006} L. Xie \textit{et al.}, Europhys. Lett. \textbf{74}, 2 (2006).

\bibitem{Goldman2006} D. Goldman and H. Swinney, Phys. Rev. Lett. \textbf{96}, 14 (2006).

\bibitem{Biggs2007} M. J. Biggs \textit{et al.}, Granul. Matter \textbf{10}, 2 (2007).

\bibitem{Maret1990} S. Fraden and G. Maret, Phys. Rev. Lett. \textbf{65}, 4 (1990).

\bibitem{Weitz1993} D. A. Weitz and D. J. Pine, in \emph{Dynamic Light Scattering: The Method and Some Applications}, Chap. 16, edited by W. Brown (Oxford University Press, Oxford, 1993).

\bibitem{Jackson2000} R. Jackson, \emph{The Dynamics of Fluidized Particles}, Chap. 3 (Cambridge University Press, Cambridge, 2000).

\bibitem{Jackson1969} T. B. Anderson and R. Jackson, Ind. Eng. Chem. Fundam. \textbf{8}, 1 (1969).

\bibitem{Gouesbet1995} N. Letaief, C. Rozé, and G. Gouesbet, J. Phys. II \textbf{5}, 12 (1995).

\bibitem{Johnsson2000} F. Johnsson \textit{et al.}, Int. J. Multiph. Flow \textbf{26}, 4 (2000).

\bibitem{Castellanos2003} J. M. Valverde, M. A. S. Quintanilla, A. Castellanos, and P. Mills,  Phys. Rev. E. \textbf{67} (2003).

\bibitem{Castellanos2005} A. Castellanos, Adv. Phys. \textbf{54}, 4 (2005).

\bibitem{Jackson1968} T. B. Anderson and R. Jackson, Ind. Eng. Chem. Fundam. \textbf{7}, 1 (1968).

\bibitem{John1989} F. MacKintosh and S. John, Phys. Rev. B \textbf{40}, 4 (1989).

\bibitem{Frenkel2012} F. Paillusson and D. Frenkel, Phys. Rev. Lett. \textbf{109}, 20 (2012).

\bibitem{Maret1990b} D. J. Pine \textit{et al.}, in \emph{Scattering And Localization Of Classical Waves In Random Media}, Vol. 8, pp. 312–-372, edited by P. Sheng (World Scientific Publishing Co. Pte. Ltd, Singapur, 1990).

\bibitem{Berne1976} B.~J. Berne and R. Pecora, \emph{Dynamic light scattering. With applications to chemistry, biology, and physics}, Chap. 4 (John Wiley \& Sons, New York, 1976).

\bibitem{Goodman1985} J. W. Goodman, \emph{Statistical Optics}, Chap. 6 (John Wiley \& Sons, New York, 1985).

\bibitem{Durian1999} P.-A. Lemieux and D. J. Durian, J. Opt. Soc. Am. A \textbf{16}, 7 (1999).

\bibitem{Cipelletti2003} L. Cipelletti \textit{et al.}, J. Phys. Condens. Matter \textbf{15}, 1 (2003).

\bibitem{Maret1987} G. Maret and P. E. Wolf, Zeitschrift f\"ur Phys. B Condens. Matter \textbf{65} 4 (1987).

\bibitem{Durian1991} D. J. Durian, D. A. Weitz, and D. J. Pine, Science \textbf{252}, 5006 (1991).

\bibitem{Durian2001} P.-A. Lemieux and D. J. Durian, Appl. Opt. \textbf{40}, 24 (2001).

\bibitem{Pusey1979} P. N. Pusey, J. Phys. A. Math. Gen. \textbf{12}, 10 (1979).

\end{thebibliography}
\end{document}